\documentclass[11pt]{article}
\usepackage{mathtools}

\usepackage{mathtools}

\usepackage{amsmath,amssymb,amstext,amsfonts}
\usepackage{amsthm}
\usepackage{graphicx}
\usepackage{subcaption}
\usepackage{float}
\usepackage[margin=0.8 in]{geometry}
\usepackage{listings}
\usepackage{xcolor}
\usepackage{subcaption}

\usepackage{graphicx}
\usepackage[utf8]{inputenc}
\newtheorem{theorem}{Theorem}[section]
\newtheorem{lemma}[theorem]{Lemma}

\usepackage{systeme}
\usepackage{comment}
\usepackage{dsfont}

\usepackage{bbm}
\usepackage{dsfont}

\usepackage{authblk}

\newcommand{\vsp}[1]{\vspace{#1 pc} \noindent}

\newcommand{\pd}[2]    { \frac{\partial #1} {\partial #2} }

\newcommand{\pdi}[2] { {\partial_#2} #1 }

\newcommand{\bvec}[1]{\ensuremath{\boldsymbol{#1}}}
\newcommand{\abs}[1]{\left| #1 \right|}
\newcommand{\CC}{{\mathbb{C}}}
\newcommand{\RR}{{\mathbb{R}}}

\newcommand{\dotp}{\boldsymbol{\cdot}}

\newcommand{\En}{\mathcal{E}}
\newcommand{\Mo}{\mathcal{M}}
\newcommand{\var}{\text{Var}}
\newcommand{\uhat}{\hat{u}}
\newcommand{\vard}[2]{\frac{\delta #1}{\delta #2}}
\newcommand{\Ham}{\mathcal{H}}
\newcommand{\HamL}{\Ham_{\Lambda}}
\newcommand{\Hthree}{\Ham_{3}}
\newcommand{\Htwo}{\Ham_{2}}
\newcommand{\Proj}{\mathcal{P}_{\Lambda}}
\newcommand{\uL}{u_{\Lambda}}
\newcommand{\UL}{U_{\Lambda}}
\newcommand{\Gibbs}{\mathcal{G}}
\newcommand{\invtemp}{\beta}

\newcommand{\SL}{\mathbb{S}^{2\Lambda-1}}
\newcommand{\pto}{\overset{p}{\to}}
\newcommand{\dto}{\overset{d}{\to}}
\newcommand{\Zpart}{Z_{\invtemp}}
\newcommand{\Zmod}{ \tilde{\Zpart} }
\newcommand{\xc}{\xi}

\newcommand{\vnorm}[1]{\abs{#1}}
\newcommand{\xv}{\bvec{x}}
\newcommand{\yv}{\bvec{y}}
\newcommand{\Xv}{\bvec{X}}
\newcommand{\Yv}{\bvec{Y}}
\newcommand{\Bv}{\bvec{B}}
\newcommand{\xhat}{\hat{\xv}}
\newcommand{\yhat}{\hat{\yv}}
\newcommand{\Xhat}{\hat{\Xv}}
\newcommand{\Yhat}{\hat{\Yv}}
\newcommand{\Bhat}{\hat{\Bv}}

\newcommand{\Xsb}{\overline{X^2_{2 \Lambda}}}
\newcommand{\Ysb}{\overline{Y^2_{2 \Lambda}}}

\newcommand{\NN}{\mathcal{N}}
\newcommand{\Id}{\bvec{I}_{2\Lambda}}
\newcommand{\Sig}{\bvec{\Sigma}}
\newcommand{\diag}{\mathop{\mathrm{diag}}}
\newcommand{\gss}{\gamma}
\newcommand{\gssS}{\gamma_{\Sig}}
\newcommand{\EV}{\mathbb{E}}
\newcommand{\oo}{\mathcal{O}}

\newcommand{\itp}{\invtemp'}

\newcommand{\edit}[1]{#1}

\begin{document}

\title{Rigorous criteria for anomalous waves induced by abrupt depth change using truncated KdV statistical mechanics}




\author[1]{Hui Sun}
\author[2]{Nicholas~J.~Moore\thanks{nickmoore83@gmail.com}}
\affil[1]{Florida State University}
\affil[2]{United States Naval Academy}


\maketitle

\begin{abstract} 
The truncated Korteweg–De Vries (TKdV) system --- a shallow-water wave model with Hamiltonian structure that exhibits weakly turbulent dynamics --- has been found to accurately predict the anomalous wave statistics observed in recent laboratory experiments. Majda et al.~(2019) developed a TKdV statistical mechanics framework based on a mixed Gibbs measure that is supported on a surface of fixed energy (microcanonical) and takes the usual macroconical form in the Hamiltonian. This paper reports two rigorous results regarding the surface-displacement distributions implied by this ensemble, both in the limit of the cutoff wavenumber $\Lambda$ growing large. First, we prove that if the inverse temperature vanishes, microstate statistics converge to Gaussian as $\Lambda \to \infty$. Second, we prove that if nonlinearity is absent and the inverse-temperature satisfies a certain physically-motivated scaling law, then microstate statistics converge to Gaussian as $\Lambda \to \infty$. When the scaling law is not satisfied, simple numerical examples demonstrate symmetric, yet highly non-Gaussian, displacement statistics to emerge in the linear system, illustrating that nonlinearity is not a strict requirement for non-normality in the fixed-energy ensemble.
The new results, taken together, imply necessary conditions for the anomalous wave statistics observed in previous numerical studies. In particular, non-vanishing inverse temperature and either the presence of nonlinearity or the violation of the scaling law are required for displacement statistics to deviate from Gaussian.
The proof of this second theorem involves the construction of an approximating measure,  which we find also elucidates the peculiar spectral decay observed in numerical studies and may open the door for improved sampling algorithms.
\end{abstract}

\section{Introduction}

Abnormally large water waves, known variously as rogue, freak, or anomalous waves, have been the subject of intense scientific scrutiny over recent decades \cite{Haver2004, Garrett2009, Hadj2014, Toffoli2017, Dem2018, Dem2019, Karm2019, dudley2019rogue}. A recent line of experimental, numerical, and theoretical work demonstrates that anomalous waves can be triggered by abrupt changes in water depth \cite{Viotti2014, Bolles2019, Majda2019, MajdaQi2019, Machine2019, Herterich2019, Trulsen2020, MooreJNS2020, Zheng2020, qi2021anomalous, li2021surface1, li2021surface2, li2021rogue, lawrence2021statistical, mendes2021non, afzal2021propagation}. In particular, a step in bottom topography is sufficient to generate highly non-normal surface-displacement statistics in a field of randomized, unidirectional water waves \cite{Viotti2014, Bolles2019, Majda2019, MooreJNS2020}.  \edit{This arrangement offers a paradigm system: one in which the conditions are sufficient to generate anomalous waves, yet simple enough to offer the hope for rigorous analysis.}

	In particular, Majda et al.~(2019) \cite{Majda2019} created a statistical mechanics framework  to help explain the experimental observations made by Bolles et al.~(2019) \cite{Bolles2019}. The framework is based on statistical and dynamical analysis of the truncated Korteweg-de Vries (TKdV) equations, with exploitation of the Hamiltonian structure to construct a particular invariant measure that is consistent with physical constraints. The invariant measure corresponds to a {\em mixed} Gibbs ensemble that is macrocanonical in Hamiltonian and {\em microcanonical in energy} \cite{Abramov2003, Majda2019, MooreJNS2020} (see Eq.~\eqref{Gibbs1} of the current paper). The latter condition mitigates the far-field, sign-indefinite divergence of the Hamiltonian, as recognized by Abromov et al.~(2003) \cite{Abramov2003} for the truncated Burgers-Hopf system. As reported in \cite{Majda2019, MajdaQi2019, MooreJNS2020}, the TKdV model accurately predicts the anomalous behavior observed in experiments, including the heightened skewness of the outgoing wave-field, elevated levels of extreme events and intermittency, the amplification of high frequencies in the spectrum, and even the detailed shape of the outgoing surface-displacement distribution. 
Other studies have demonstrated the ability to predict these, and related, extreme events through machine learning or stochastic parameterization \cite{Machine2019, Guth2019, Chen2020}.

	The above studies provide ample numerical evidence for the creation of anomalous wave statistics within the TKdV framework \cite{Majda2019, MajdaQi2019, MooreJNS2020} and offer a relatively transparent statistical description in terms of the system's Hamiltonian, which is a macrostate quantity.
\edit{
The simplicity of this statistical description offers several advantages, for example: 
(1) it illustrates how the depth change enhances nonlinearity and thus promotes positive skewness in the outgoing wave field \cite{Majda2019, MooreJNS2020};  (2) it leads to an explicit formula linking the outgoing surface-displacement skewness to the change in slope variance. This formula, in fact, has been verified by experimental measurements \cite{MooreJNS2020}.}
However, the detailed {\em distributions of surface displacements}, or the microstates, implied by the theory have remained more elusive. So far, these microstate distributions have required  numerical computation from either dynamical simulations or numerical sampling of the Gibbs ensemble. It is precisely these surface-displacement distributions that are of direct interest to practitioners who seek to  quantify and predict anomalous waves in the field. 

	This paper takes a major step towards providing the precise, mathematical link between the Gibbs-based macrostate description and the microstate statistics implied by it.
\edit{Our main focus is to determine which conditions lead to Gaussian statistics of the surface displacement and which may result in anomalous statistics.} We first prove that the zero inverse-temperature case, $\invtemp=0$, of the Majda's mixed Gibbs ensemble implies Gaussian surface-displacement statistics in the limit of the cutoff wavenumber, $\Lambda$, growing large. 
\edit{The proof is elementary and it provides the foundation for subsequent analysis. Second, we examine the special case of positive inverse temperature, $\invtemp>0$, with vanishing nonlinearity; i.e.~the linear TKdV system, also known as the (truncated) Airy equation. From the form of the Hamiltonian, it is easy to see that nonlinearity promotes skewness in surface-displacement statistics. Furthermore, with nonlinearity absent, one can show that the surface-displacement distributions must be symmetric. However, for certain values of $\Lambda$ and $\invtemp$, simple numerical sampling experiments demonstrate that the linear TKdV system can produce displacement statistics that, while they remain symmetric, are strongly non-Gaussian and exhibit significant kurtosis.
Laboratory experiments often observe near Gaussian statistics when nonlinearity is sufficiently small, thus raising the question of what conditions are needed to restore normality. We resolve this question by identifying a scaling relationship between the inverse temperature, the cutoff wavenumber, and the dispersive coefficient that guarantees convergence to Gaussian statistics in the limit $\Lambda \to \infty$. This proof is more involved than the $\beta=0$ case and requires the construction of a particular approximating measure. In addition to its value in the proof, we find this approximating measure lends additional physical insight into the TKdV system, as it elucidates a peculiar spectral decay that has been observed in previous numerical studies \cite{Majda2019, MajdaQi2019, MooreJNS2020}. Taken together, our rigorous results outline sufficient conditions for Gaussian statistics or, equivalently, necessary conditions for anomalous statistics. In particular, for surface-displacement statistics to deviate from Gaussian, the inverse temperature must be non-zero {\em and} either nonlinearity must be present or the scaling-law must be violated.
}

	Related studies have performed statistical analysis on the nonlinear Schr{\"o}dinger equation \cite{Lebowitz1988, Lebowitz1989, Bourgain1994, Bourgain1999}, a partial differential equation (PDE) with Hamiltonian structure similar to KdV. In particular, both PDEs exhibit the same far-field, sign-indefinite divergence of the simple canonical measure. To avoid this divergence, many studies impose an upper bound on energy, $\En < \En_0$, which offers a considerable simplification in that one component of the invariant measure reduces to Weiner measure \cite{Bourgain1994, Bourgain1999}.
By contrast, the mixed microcanonical-macrocanonical ensemble of Majda et al.~(2019) \cite{Majda2019} --- whose validity has been borne out by careful comparison to laboratory experiments --- fixes the value of energy, $\En = \En_0$. Geometrically, dynamics are confined to the {\em surface} of a hypersphere in spectral space, rather than its interior. 
\edit{Other studies have explicitly recognized this fixed-energy ensemble as the `physically preferred choice' \cite{Lebowitz1988, Lebowitz1989}, but choose the bounded-energy formalism for technical convenience.
In the case of fixed-energy, we will show that the analogous component of the mixed Gibbs measure does {\em not}  reduce to Weiner measure and exhibits a markedly different spectral structure. We therefore cannot rely on the simple $1/k$ spectral decay of Weiner measure \cite{Bourgain1994, Bourgain1999}, nor do we artificially impose an empirical spectrum with randomly sampled multipliers \cite{Dem2018, Dem2019}. Instead, we construct a non-trivial measure that converges to the Gibbs measure of the linear system as $\Lambda \to \infty$, and the corresponding spectrum is allowed to emerge naturally. Thus, rather than being rooted in an empirical JONSWAP spectrum, our study firmly links the observable statistics to the fundamental physical principles of an entropy-maximizing Gibbs ensemble.}

\edit{
Interestingly, we find that this approximating measure, initially constructed as a theoretical tool, also appears to capture the spectral decay of the {\em nonlinear} system fairly well. This observation may prove valuable for future importance-sampling algorithms, as discussed further in Section \ref{ImportanceSampling}.}

There exist wide range of complex geophysical problems for which the extraction of statistical features is a driving interest. Examples abound from climate science \cite{RothPNAS2019, Caves2019, Rothman2017}, atmospheric science \cite{OgStChMajda2019, Yang2019, Chen2018, StechHot2016}, morphology formation \cite{mac2022morphological, Huang2020, Chiu2020, Quaife2018, RothProcA2017, McDonald2020}, and thermal convection \cite{Han2020, McCurdy2019, Mac2018, zhang2000periodic, liu2008self, zhang1997non}. The statistical analysis presented in this paper may ultimately prove valuable for such applications.

The outline of the paper is follows. In Section \ref{Background}, we provide the relevant physical and mathematical background, including the Hamiltonian structure of TKdV and the fixed-energy, mixed Gibbs ensemble. In Section \ref{Results}, we report the main results of the paper, including the numerical experiments that demonstrate nonlinearity is not a strict requirement for non-Gaussian statistics, and Theorems \ref{thm1} and \ref{thm2}, each of which outline conditions that guarantee  convergence to Gaussian statistics.
Central to the proof of Theorem \ref{thm2} is the construction of the approximating measure mentioned above. We discuss some salient properties of this measure in Section \ref{ImportanceSampling} and provide concluding remarks in Section \ref{Conclusion}.


\section{Physical background and mathematical formulation}
\label{Background}

\subsection{The KdV system and Hamiltonian structure}

	The laboratory experiments of Bolles et al.~(2019) \cite{Bolles2019} provide the physical motivation for our study. In these experiments, a randomized field of unidirectional water waves propagates through a region of constant depth, encounters an abrupt depth change (ADC) in the bottom topography, and then continues into a region of shallower depth. Bolles et al.~(2019) \cite{Bolles2019} discovered that an initial Gaussian distribution of surface waves can become highly skewed upon encountering the ADC, with an elevated level of extreme events and enhanced intermittency a short distance downstream. Inspired by these observations, Majda et al.~(2019) \cite{Majda2019} developed the statistical mechanical TKdV framework discussed above and demonstrated its ability to recover a wide range of the experimental measurements \cite{Majda2019, MooreJNS2020}.

	The theory exploits the Hamiltonian structure of the TKdV system to construct invariant Gibbs measures that describe wave statistics upstream and downstream of the ADC. Importantly, the upstream and downstream measures are furnished with distinct inverse temperatures. This theory views the upstream state as an incoming wave-field with inverse temperature that is set externally, for example by the experimental apparatus or, in the ocean, by the strength and character of the wind or tidal forcing. The downstream state, however, is slave to the upstream one as determined by a statistical matching condition enforced at the ADC \cite{Majda2019, MooreJNS2020}. Ultimately, this matching condition yields the outgoing inverse temperature as a function of the specified incoming inverse temperature. By modifying the coefficients that enter the Hamiltonian and thus the Gibbs measure, the depth change can dramatically alter the character of the randomized waves and produce anomalous statistics.

	The present study limits attention to the outgoing dynamics, supposing that its inverse temperature has already been set by enforcing the statistical matching condition. Our starting point is therefore the {\em constant-depth} KdV equation \cite{Lax1975, Johnson1997, Whitham2011} for surface displacement $u$ depending on horizontal location $\xc$ and time $t$:
\begin{align}
\label{KdV}
&u_t + C_3 \, u u_\xc + C_2 \, u_{\xc \xc \xc} = 0
\qquad \text{for } \xc \in [-\pi,\pi]
\end{align}		
This equation holds in a {\em moving} reference frame that travels with the leading-order wave speed. Boundary conditions are periodic on the normalized domain $\xc \in [-\pi,\pi]$. All variables $u, \xc, t$ are assumed to be dimensionless already, and the dimensionless coefficients $C_3$ and $C_2$ characterize the strength of nonlinearity and dispersion respectively (see Moore et al.~2020 \cite{MooreJNS2020} for how these dimensionless parameters relate to physical ones). The effect of the depth change is to increase the value of $C_3$ and decrease that of $C_2$, thereby simultaneously promoting nonlinearity and suppressing dispersion. Once again, the present study limits attention to the downstream state, so that $C_3$ and $C_2$ remain constant.

	The KdV equation \eqref{KdV} enjoys a Hamiltonian structure, most easily defined by introducing the components
\begin{align}
\label{H3H2}
\Hthree = \frac{1}{6} \int_{-\pi}^{\pi} u^3 \, d\xc	\, , \qquad
\Htwo = \frac{1}{2} \int_{-\pi}^{\pi} \left( \pd{u}{\xc} \right)^2 \, d\xc	\, .
\end{align}
Then the Hamiltonian can be expressed as
\begin{equation}
\label{KdVHam}
\Ham = C_2 \, \Htwo - C_3 \, \Hthree
\end{equation}
This expression makes clear the notational choices for the coefficients $C_2$ and $C_3$ above. The KdV equation \eqref{KdV} can then be written as
\begin{align}
\label{HamStruct}
\pd{u}{t} = \pd{}{\xc} \vard{\Ham}{u}
\end{align}
where $\pdi{}{\xc}$ is a symplectic operator. Hence KdV \eqref{KdV} is a Hamiltonian system and, consequently, the Hamiltonian \eqref{KdVHam} is conserved during evolution.

In addition, momentum and energy are conserved under KdV dynamics
\begin{align}
\label{MomEn}
\Mo[u] \equiv \int_{-\pi}^{\pi} u \, d\xc \, = 0 , \qquad
\En[u] \equiv \frac{1}{\pi} \int_{-\pi}^{\pi} u^2 \, d\xc = 1
\end{align}
The momentum vanishes, as indicated above, because $u$ is measured as displacement from equilibrium. Meanwhile, the energy has been normalized to unity due to the choice of characteristic wave amplitude. We remark that, compared to Moore et al.~(2020) \cite{MooreJNS2020}, we have rescaled $\En$ by a factor of $2/\pi$ to simplify the forthcoming analysis.

\subsection{The truncated KdV system and Hamiltonian structure}

	Following Majda et al.~(2019) \cite{Majda2019}, we perform a finite Galerkin truncation of \eqref{KdV}. To this end, consider a spatial Fourier representation of the state variable 
\begin{align}
&u(\xc,t) = \sum_{k=-\infty}^{\infty} \uhat_k(t) \, e^{i k \xc} 
=  \sum_{k=1}^{\infty} a_k(t) \cos(k \xc) + b_k(t) \sin(k \xc) \, , \\
\label{uhat}
&\uhat_k = \frac{1}{2} (a_k - i b_k)= \frac{1}{2 \pi} \int_{-\pi}^{\pi} u(\xc,t) \, e^{-i k \xc} \, d\xc \, .
\end{align}
For convenience, we have recorded both the real and the complex Fourier representations, $\uhat_k \in \CC$ and $a_k, b_k \in \RR$. Hereafter, we will usually suppress the time dependence of these coefficients. Note that $\uhat_{-k} = \uhat_{k}^*$ since $u(\xc,t)$ is real valued and $\uhat_0 = 0$ due to momentum vanishing.
Next, consider the Galerkin truncation at wavenumber $\Lambda$
\begin{align}
\label{uL}
\uL(\xc,t) = \Proj u = \sum_{\abs{k} \le \Lambda} \uhat_k \, e^{i k \xc} 
= \sum_{k=1}^{\Lambda} a_k \cos(k \xc) + b_k \sin(k \xc) \, ,
\end{align}
where $\Proj$ is a projection operator and \eqref{uhat} still holds. Inserting the projected variable, $\uL$, into the KdV equation and applying the projection operator, $\Proj$, again where necessary produces the truncated KdV equation (TKdV) \cite{Bajars2013, Majda2019, MooreJNS2020}
\begin{align}
\label{TKdV}
&\pd{\uL}{t} + \frac{1}{2} C_3 \, \pd{}{\xc} \Proj (\uL)^2 
+ C_2 \, \frac{\partial^3 \uL}{\partial \xc^3} = 0
\qquad \text{for } \xc \in [-\pi,\pi]
\end{align}
Equation \eqref{TKdV} represents a {\em finite}-dimensional dynamical system. The quadratic nonlinearity, $\pdi{}{\xc}  \Proj (\uL)^2$, mixes the modes during evolution, and the additional projection operator in this term removes the aliased modes of wavenumber larger than $\Lambda$. Typical values of the cutoff wavenumber used in the previous studies are $\Lambda = $ 8--32 \cite{Majda2019, MooreJNS2020}, whereas Majda \& Qi (2019) found exact solutions for more severe truncations, as low as $\Lambda=2$ \cite{MajdaQi2019}. The rigorous results obtained in the current study, however, hold in the limit of large cutoff-wavenumber, $\Lambda \to \infty$. We remark that this analysis is not necessarily the same as direct analysis of the continuous KdV system  \eqref{KdV}.

The TKdV equation \eqref{TKdV} enjoys nearly the same Hamiltonian structure as KdV, with the only modification being the inclusion of the projection operator,
\begin{align}
\label{TKdVHam}
&\HamL = C_2 \, \Htwo[\uL] - C_3 \, \Hthree[\uL] \, , \\
&\pd{}{t} {\uL} = \pd{}{\xc} \Proj \, \vard{\HamL}{\uL}
\end{align}
where now $\pdi{}{\xc} \Proj$ is the symplectic operator of interest. 

	The system's microstate can either by described in physical space $\uL(\xc, t)$, in complex spectral space, $(\uhat_1, \uhat_2, \cdots, \uhat_{\Lambda}) \in \CC^{\Lambda}$, or in real spectral space $(a_1, a_2, \cdots, a_{\Lambda}, b_1, b_2, \cdots, b_{\Lambda}) \in \RR^{2 \Lambda}$. All are equivalent through \eqref{uhat}--\eqref{uL}, and this paper primarily uses the real spectral representation.
The momentum and energy defined in \eqref{MomEn} are also conserved in the truncated system and have the same normalized values $\Mo[\uL] = 0$ and $\En[\uL] = 1$. Parseval's identity implies
\begin{equation}
\label{Econd}
\En[\uL] = 4 \sum_{k=1}^{\Lambda} \abs{\uhat_k}^2 = 
\sum_{k=1}^{\Lambda} a_k^2 + b_k^2 =1
\end{equation}
Thus, in real spectral space, the TKdV dynamics of interest $\xv(t) := (a_1(t), \cdots, a_{\Lambda}(t), b_1(t), \cdots, b_{\Lambda}(t))$
are confined to the unit hypersphere, $\SL = \{\xv \in \RR^{2 \Lambda}: \vnorm{\xv} = 1\}$, which is a compact set. This geometric interpretation is of central importance to the present study.

\subsection{The mixed canonical-microcanonical Gibbs ensemble}

	Following Majda et al.~(2019) \cite{Majda2019}, we employ a {\em mixed canonical-microcanonical} Gibbs measure $\Gibbs$ to examine the statistical mechanics of \eqref{TKdV}. Similar to that used in the truncated Burgers-Hopf system \cite{majda2000remarkable, majda2002statistical, Abramov2003}, this ensemble is microcanonical in energy and canonical in the Hamiltonian. The basic idea can be gleaned through the abstract representation \cite{Bajars2013}
\begin{align}
\label{Gibbs1}
d \Gibbs \propto \exp(-\invtemp \Ham) \delta(\En - 1)
\end{align}
where $\invtemp$ is the system's inverse temperature. The exponential dependence with respect to the Hamiltonian is the well-known canonical distribution, which, under suitable conditions, maximizes entropy \cite{MajdaWang2006}. The role of the Dirac-delta term $\delta(\En - 1)$ is to confine the distribution to the compact set $\En = 1$, thus avoiding the far-field, sign-indefinite divergence of the cubic component $\Hthree$ and thereby creating a normalizable distribution \cite{Abramov2003, Bajars2013, Majda2019, MooreJNS2020}. Previous studies have found that a positive inverse temperature, $\invtemp > 0$, yields a physically realistic decaying spectrum that agrees with laboratory experiments \cite{Majda2019, MooreJNS2020}. We remark that, when equipped with measure \eqref{Gibbs1}, the TKdV dynamical system is not strictly ergodic, but numerical evidence suggests that weak thermostatting of just the largest mode is sufficient to produce ergodic properties \cite{Bajars2013}.

	To gain a practical understanding of the mixed Gibbs measure \eqref{Gibbs1}, first consider the special case of zero inverse temperature, $\invtemp = 0$, for which  \eqref{Gibbs1} reduces to the {\em uniform measure} on $\SL$. The uniform measure can be identified with integration over $\RR^{2\Lambda}$ by sampling from any rotationally invariant distribution and then normalizing to the unit hypersphere $\SL$ \cite{Abramov2003}. More specifically, consider a random vector $\Xv = (X_1, X_2, \cdots, X_{2\Lambda}) \in \RR^{2\Lambda}$ drawn from any rotationally invariant distribution on $\RR^{2\Lambda}$. Identify this vector with a specific microstate by defining the real coefficients,
\begin{equation}
\label{abdefn}
a_k = X_k / \vnorm{\Xv} \, , \quad
b_k = X_{\Lambda+k} / \vnorm{\Xv} \, , \quad \text{for } k = 1, 2, \cdots \Lambda
\end{equation}
where $\vnorm{\Xv} = \left( \sum_{k=1}^{2 \Lambda} X^2_k \right)^{1/2}$ is the standard 2-norm. This normalization guarantees that \eqref{Econd} is satisifed, and the microstates corresponding to \eqref{abdefn} are uniformly distributed on $\SL$ \cite{Abramov2003}.

	For simplicity, we use the standard normal distribution on $\RR^{2\Lambda}$ as the rotationally invariant measure:
\begin{equation}
\label{stand_norm}
d \gss(\xv) :=  \prod_{k=1}^{2\Lambda} \frac{1}{\sqrt{2 \pi}} e^{ {-x_k^2}/{2} } \, dx_k
\end{equation}
Then, for any measurable function on the unit hypersphere $\phi: \SL \to \RR$, integration with respect to the uniform measure $d\Gibbs_0$ on $\SL$ can be written as
\begin{equation}
\label{Gibbs0}
\int_{\SL} \phi \, d\Gibbs_0 = \int_{\RR^{2 \Lambda}} \phi ( \xhat ) \, d\gss(\xv)
\end{equation}
where $\xhat = \xv/\vnorm{\xv}$ is a unit vector.

	We can now precisely define the Gibbs measure \eqref{Gibbs1} for arbitrary $\invtemp$. For a fixed location $\xc$ and time $t$, and for any measurable function $\phi: \SL \to \RR$, let
\begin{equation}
\label{Gibbs2}
\int_{\SL} \phi \, d\Gibbs = \Zpart^{-1} \int_{\RR^{2 \Lambda}} \phi ( \xhat ) \,
\exp( -\invtemp \, \Ham[ \UL( \xhat ) ] ) \, d\gss(\xv)
\end{equation}
\edit{Above, $\UL(\xhat) := \uL(\xc, t)$ identifies a unit vector $\xhat \in \SL$ with a particular microstate $\uL$ through \eqref{abdefn} and \eqref{uL}. }Meanwhile, $\Zpart$ is the partition function, i.e.~a normalization constant that depends on $\invtemp$ and $\Lambda$.

The analysis in this paper relies only on the mixed Gibbs ensemble  \eqref{Gibbs2}, and not on any dynamics. Hence, in the remainder of the paper, we fix an arbitrary location in space $\xc$ and time $t$, and analyze the corresponding microstates $\uL(\xc,t)$ sampled from \eqref{Gibbs2}. We first establish a simple Lemma that will be helpful in the analysis:
\begin{lemma}[Spatiotemporal Independence]
\label{space_ind}
Microstate statistics of $\uL(\xc, t)$ corresponding to measure \eqref{Gibbs2} are independent of location $\xc$ and time $t$.
\end{lemma}
\begin{proof}[{\bf Proof}]
The temporal independence is trivial since the Hamiltonian  \eqref{TKdVHam} is independent of time.
Spatial independence follows from the fact that translation of the periodic function $\uL(\xc,t)$ in $\xc$ preserves both components $\Htwo$ and $\Hthree$, given in Eq.~\eqref{H3H2}, and therefore preserves the Hamiltionian that is defined in Eq.~\eqref{TKdVHam} and used in the Gibbs measure \eqref{Gibbs2}.
\end{proof}

\section{Results}
\label{Results}

In this section, we report the two main theorems of the paper. Both concern the statistical distributions of surface displacement, $\uL$, implied by the mixed Gibbs ensemble \eqref{Gibbs2} with inverse temperature $\invtemp$. Once again, we have fixed an arbitrary position in space $\xc$ and time $t$, with the aid of Lemma \ref{space_ind}, and so we have suppressed the dependence of $\uL$ on these variables.
The first theorem treats the case of $\invtemp=0$ with arbitrary  $C_3$ and $C_2$, while the second treats the case of $\invtemp >0$ with nonlinearity absent, $C_3 = 0$. 

\subsection{TKdV with zero inverse temperature}

We now introduce the first theorem:
\begin{theorem}[Zero Inverse Temperature]
\label{thm1}
\edit{Consider the TKdV mixed-Gibbs ensemble \eqref{Gibbs2} with zero inverse temperature, $\invtemp = 0$, and arbitrary choices of $C_2$ and $C_3$. Under these conditions, the surface displacement, $\uL$, convergences in probability to a normal random variable with mean zero and variance $1/2$.}
\end{theorem}

\begin{proof}[{\bf Proof}]

Consider the vector of the real Fourier basis elements evaluated at location $\xc$
\begin{equation}
\label{Bv}
\Bv := \left( \cos \xc, \cos 2\xc, \cdots, \cos \Lambda \xc, \sin \xc, \sin 2\xc, \cdots, \sin \Lambda \xc \right) \in \RR^{2\Lambda}
\end{equation}
As suggested by \eqref{stand_norm}, let $\Xv = (X_1, X_2, \cdots, X_{2\Lambda}) \in \RR^{2\Lambda}$ be a standard normal random vector $\Xv \sim \NN(0,\Id)$ (i.e.~the components $X_k$ are i.i.d.~with $X_k \sim \NN(0,1)$), and let $\Xhat = \Xv / \vnorm{\Xv}$ be the corresponding unit vector.
Then the corresponding microstate $\uL(\xc,t) = \UL(\Xhat)$ defined by \eqref{abdefn} can be written as
\begin{equation}
\UL(\Xhat) = \Bv \dotp \Xhat
\end{equation}
where $\dotp$ is the standard dot product on $\RR^{2\Lambda}$.
This is nothing more than a vector representation of the real Fourier series \eqref{uL}, with standard normal i.i.d.~coefficients, $a_k$ and $b_k$, that have been normalized to satisfy the unit-energy constraint \eqref{Econd}. Let $\Bhat = \Bv/ \vnorm{\Bv}$ and note that $\vnorm{\Bv} = \sqrt{\Lambda}$. Then the above can be rewritten as
\begin{equation}
\label{uBX}
\UL = \frac{\vnorm{\Bv}}{\vnorm{\Xv}} \,\,  \Bhat \dotp \Xv
= \frac{\sqrt{\Lambda}}{\vnorm{\Xv}} \,\,  \Bhat \dotp \Xv
\end{equation}
Consider the scalar random variable 
\begin{equation}
\frac{\vnorm{\Xv}^2}{2 \Lambda} 
=  \frac{1}{2 \Lambda} \sum_{k=1}^{2 \Lambda} X^2_k 
\end{equation}
where the components $X_k$ are i.i.d.~with $X_k \sim \NN(0,1)$.
By the law of large numbers (LLN),
\begin{equation}
\label{LLN1}
\frac{\vnorm{\Xv}^2}{2 \Lambda} \pto 1
\qquad \text{as } \Lambda \to \infty
\end{equation}
where $\pto$ indicates convergence in probability \cite{Varadhan2001}.

Now, returning to \eqref{uBX}, the scalar product $\Bhat \dotp \Xv$ is a finite sum of independent normal random variables, hence is a normal random variable. Furthermore, since $\Bhat$ is a unit vector, the variance is unity, $\Bhat \dotp \Xv \sim \NN(0,1)$. Substituting the limiting value \eqref{LLN1} into \eqref{uBX} implies that, in the limit $\Lambda \to \infty$, $\uL$ converges to a normal random variable with variance $1/2$:
\begin{equation}
\edit{\uL \pto \NN(0, 1/2)}
\quad \text{as } \Lambda \to \infty
\end{equation}
\end{proof}

\subsection{Linear TKdV with positive inverse temperature}

	Theorem \ref{thm1} indicates that zero inverse temperature, $\invtemp = 0$, leads to Gaussian surface-displacement statistics, consistent with previous numerical results \cite{Majda2019, MooreJNS2020}.
\edit{
Those studies found that positive inverse temperatures, $\invtemp > 0$, can produce non-Gaussian displacement statistics, often with a high degree of skewness and sometimes even a bimodal structure \cite{MajdaQi2019}. Through the Hamiltonian, it is easy to see that nonlinearity, represented by the component $\Hthree$ in \eqref{H3H2}, promotes skewness in Gibbs ensemble \eqref{Gibbs1}, and is therefore at least partially responsible for the non-Gaussian statistics. This observation raises the question of whether nonlinearity is strictly required for non-Gaussian statistics. In short, we find the answer to be {\em no}, with some simple numerical experiments demonstrating  strongly non-Gaussian features to arise in the linear TKdV Gibbs ensemble provided that the inverse temperature is sufficiently large. In particular, Fig.~\ref{FigHists} shows three histograms of the surface displacement, $\uL$, numerically sampled from \eqref{Gibbs2} with $C_3=0$. Each of these histograms exhibits a significant departure from Gaussian --- either enhanced flatness near the middle as in Fig.~\ref{FigHists}(b) or bimodality as in Fig.~\ref{FigHists}(a) and (c) --- despite nonlinearity being completely absent. 

Figure \ref{FigHists}(a) demonstrates non-Gaussian features  under the simplest possible circumstances of a single complex mode $\Lambda=1$. For this case, a simple thought experiment is enough to realize that ensemble statistics cannot be Gaussian. Indeed, since the energy is fixed, $\En = 1$, in Gibbs ensemble \eqref{Gibbs1}, all sampled microstates are unimodal sine-waves of the {\em same} amplitude (though different phases). Therefore, the ensemble statistics follow the so-called arcsine distribution, characterized by sharp peaks at the extrema as evident in Fig.~\ref{FigHists}(a). This situation contrasts with the simpler, bounded-energy construct $\En < \En_0$, for which the random amplitudes of sampled microstates permit near-Gaussian ensemble statistics even for $\Lambda=1$. Furthermore, this example illustrates the subtlety of the fixed-energy ensemble --- a construct that other authors have explicitly recognized as the `physically preferred choice', but instead chose the bounded energy ensemble for technical convenience \cite{Lebowitz1988, Lebowitz1989}. Here, we aim to confront some of the technical challenges presented by the the fixed-energy ensemble, rather than avoid them.



In this first example with $\Lambda=1$, the Gibbs ensemble \eqref{Gibbs2} is independent of the inverse temperature $\invtemp$, since each microstate consists of a single mode whose magnitude becomes normalized to satisfy $\En=1$. For $\Lambda>1$, however, the inverse temperature plays a strong role. We introduce a normalized inverse temperature,
\begin{equation}
\itp = \pi \invtemp C_2 \Lambda^2 /2 \, ,
\end{equation}
which will prove convenient in the theoretical analysis.

Figure~\ref{FigHists}(b)--(c) shows microstate histograms for the case $\Lambda=4$ and two choices of $\itp$. These figures show non-Gaussian features to persist even when the number of modes is increased.  As seen in Fig.~\ref{FigHists}(b), $\itp = 10$ produces a visibly flat distribution as compared to a Gaussian of the same variance (green dotted curve). The excess kurtosis is -0.7, indicating a significant departure from Gaussian statistics. Meanwhile, Fig.~\ref{FigHists}(c) shows that increasing $\itp$ to 50 causes the bimodal structure seen previously to reemerge. This observation sheds light on previous numerical simulations, which showed skewed, bimodal distributions to arise dynamically under conditions of strong nonlinearity and large inverse temperature \cite{MajdaQi2019}. Such distributions were associated with phase-change behavior, but little explanation was given for why they emerge, other than observing their formation in direct numerical simulations. Our simple, sampling experiments suggest such states result from the combination of two separate effects: bimodality results from a large ratio of inverse temperature to cutoff wavenumber (i.e.~large $\itp$), while  skewness results from strong nonlinearity as evident from the Hamiltonian structure.

\begin{figure}
\begin{center}
\includegraphics[width = 0.99 \linewidth]{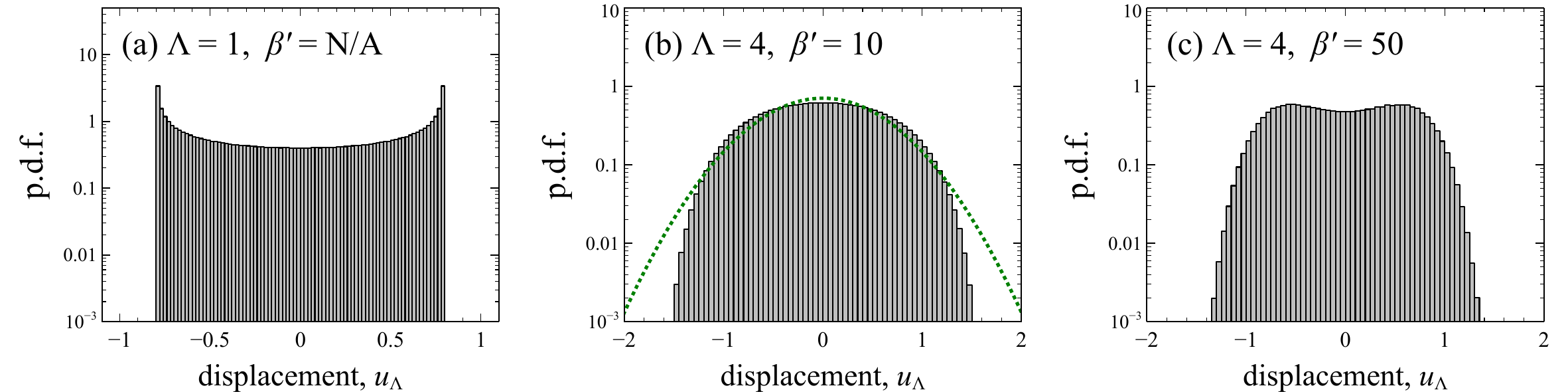}
\caption{
\edit{Non-Gaussian statistics in linear TKdV: Numerical experiments demonstrate strongly non-Gaussian features in microstates sampled from Gibbs measure \eqref{Gibbs2}, despite nonlinearity being completely absent. (a) For a single mode, $\Lambda=1$, statistics follow the arcsine distribution. (b) For $\Lambda=4$ and $\itp=10$, the distribution is visible flat compared to a Gaussian of the same variance (green dotted curve). (c) For higher $\itp$, the bimodality reemerges.}
}
\label{FigHists}
\end{center}
\end{figure}

Experimental studies, meanwhile, have typically observed near Gaussian statistics to arise under conditions of sufficiently weak nonlinearity, raising the question of how to reconcile the above numerical examples. 
The resolution is that all of these examples feature a moderate or small number of modes and a relatively large inverse temperature, giving $\itp \gg 1$. 
Below, we will prove that if the scaling law 
\begin{equation}
\label{scaling}
\itp = \oo(1) \qquad \text{for } \Lambda \gg 1
\end{equation}
is satisfied, then ensemble statistics of the {\em linear} TKdV system converge to Gaussian in the limit of large $\Lambda$. In terms of primitive quantities, this scaling law is 
$\invtemp C_2 = \oo(\Lambda^{-2})$, which has two possible physical interpretations.}
First, one can consider fixed $\invtemp$ and  $C_2 \sim \Lambda^{-2}$, which is a scaling that was justified on physical grounds by Moore et al.~(2020)  \cite{MooreJNS2020} and was found to produce consistency with laboratory measurements. In essence, this scaling requires the dispersive coefficient to decay with the cutoff wavenumber so that the peak spectral frequency is resolved in the TKdV model and, furthermore, lies near the logarithmic center of the resolved modes. Alternatively, one could consider $C_2$ fixed and $\invtemp \sim \Lambda^{-2}$, which would be the interpretation most appropriate for relating to the continuous KdV system \eqref{KdV} with a non-vanishing dispersive coefficient $C_2$. It is interesting, and perhaps meaningful, that our results suggest a particular scaling of the inverse temperature in order to make that connection. Interpretation aside, we henceforth assume that the scaling relation \eqref{scaling} holds to prove converge to Gaussian statistics in the linear system.

For linear TKdV ($C_3=0$) the Hamiltonian reduces to
\begin{equation}
\label{linHam}
\Ham
= \frac{1}{2} C_2 \int_{-\pi}^{\pi} \left( \pd{u}{\xc} \right)^2 \, d\xc
= \frac{\pi C_2}{2  \vnorm{\xv}^2}  \sum_{k=1}^{\Lambda} k^2 \left(x_k^2 + x_{\Lambda+k}^2 \right)
\end{equation}
where we have used Parseval's identify. \edit{Below, we will prove the surface displacement, $\uL(\xc,t) = \UL(\xhat)$, converges in distribution to a normal random variable.} 
That is, for any bounded, continuous function $f \in C_b(\RR \to \RR)$, we will show $\EV f(\uL) \to \EV f(Z)$ as $\Lambda \to \infty$, where $Z$ is a normal random variable. Computing the expected value of such a function $f \in C_b(\RR \to \RR)$ gives
\begin{equation}
\label{LinGibbs}
\EV f(\uL) = \int_{\SL} f(\UL(\xhat)) \, d\Gibbs
= \Zpart^{-1} \int_{\RR^{2 \Lambda}} f(\UL(\xhat)) \,
\exp \left( - \frac{ \itp}{ \Lambda^2 \vnorm{\xv}^2 } 
\sum_{k=1}^{\Lambda} k^2 \left(x_k^2 + x_{\Lambda+k}^2 \right) \right) \, d\gss(\xv)
\end{equation}
\edit{We remark that, with the above formula, it is easy to see that the linear system always produces a symmetric surface-displacement distribution, regardless of the value of $\itp$. Indeed, the one-to-one mapping $\xv \to -\xv$ preserves the Hamiltonian \eqref{linHam} of linear TKdV, making any displacement value $\uL$ and its negative $-\uL$ equally likely. This observation is confirmed by the numerical examples from Fig.~\ref{FigHists}, which though strongly non-Gaussian, are clearly symmetric distributions. We next prove that if scaling law \eqref{scaling} is satisfied, not only is the surface-displacement distribution symmetric,  but it also converges to a normal distribution.
}

\begin{theorem}[Linear TKdV]
\label{thm2}
\edit{
Consider the TKdV mixed-Gibbs ensemble \eqref{Gibbs2} with nonlinearity absent, $C_3=0$, with positive inverse temperature $\invtemp > 0$, and with scaling law \eqref{scaling} satisifed. Under these conditions, the surface displacement, $\uL$, converges in distribution to a normal random variable with mean zero and variance $1/2$.}
\end{theorem}

\begin{proof}[{\bf Intuitive argument (non-rigorous)}]

The essence of the proof is fairly intuitive, and so we provide this main part of the argument before going back to fill in some technical details. As seen in the proof of Theorem \ref{thm1}, for a standard normal vector $\Xv \sim \NN(0, \Id)$, the LLN gives $\vnorm{\Xv}^2/(2 \Lambda) \pto 1$  as $\Lambda \to \infty$. This relationship motivates the naive substitution $\vnorm{\xv}^2 \approx 2\Lambda$ in \eqref{LinGibbs}, or more generally $\vnorm{\xv}^2 \approx 2\Lambda / \alpha$, where $\alpha > 0$ is simply an extra degree of freedom. After making this substitution, some straightforward calculation produces
\begin{align}
\label{LinGibbsApprox}
\EV f(\uL) \approx \int_{\RR^{2 \Lambda}} f(\UL( \yhat )) \, d \gssS(\yv)
\end{align}
where $\gssS$ is a modified Gaussian measure with covariance matrix
\begin{align}
\label{Sig}
&\Sig = \diag \left(\sigma_1^2, \sigma_2^2, \cdots, \sigma_{\Lambda}^2, \sigma_1^2, \sigma_2^2, \cdots, \sigma_{\Lambda}^2 \right) \\
\label{sigma}
&\sigma_k^2 = \frac{1}{1 + \alpha \itp k^2/\Lambda^3}
\qquad \text{for } k=1,2,\cdots, \Lambda
\end{align}
More explicitly
\begin{equation}
\label{dgSig}
d \gssS(\yv) = \prod_{k=1}^{\Lambda} \frac{1}{{2 \pi \sigma_k^2}} 
\exp \left( -\frac{y_k^2 + y_{\Lambda+k}^2} {2 \sigma_k^2}\right) dy_k dy_{\Lambda+k}
\end{equation}
The above is a change of measure from the standard, isotropic Gaussian $d\gss$ with covariance $\Id$, to an anisotropic Gaussian with covariance $\Sig$. Hence, we can follow the same reasoning as in the proof of Theorem \ref{thm1}, except with a random vector $\Yv \in \RR^{2\Lambda}$ drawn from the altered normal distribution $\Yv \sim \NN(0, \Sig)$. Notice that in \eqref{LinGibbsApprox} the normalization constant $\Zpart$ has dropped out, which can be seen by simply taking $\phi \equiv 1$ as the test function. Also, note that $\sigma_k$ is bounded above and below
\begin{align}
\label{sigma_bound}
\frac{1}{1 + \alpha \itp/\Lambda} < \sigma_k^2 < 1
\qquad \text{for } k=1,2,\cdots, \Lambda
\end{align}

Mirroring the proof of Theorem \ref{thm1}, we fix an arbitrary location $\xc$ and consider the vector of basis functions  $\Bv \in \RR^{2\Lambda}$ from \eqref{Bv}, which has corresponding microstate
\begin{equation}
\label{uL2}
\uL = \UL(\Yhat) = \frac{\vnorm{\Bv}}{\vnorm{\Yv}} \,\,  \Bhat \dotp \Yv  = \frac{\sqrt{\Lambda}}{\vnorm{\Yv}} \,\,  \Bhat \dotp \Yv
\end{equation}
where $\Yv \sim \NN(0, \Sig)$. The scalar product $\Bhat \dotp \Yv$ is a finite sum of independent normal random variables (though not identically distributed) and hence is a normal variable \cite{Varadhan2001}.
Considering the scalar random variable 
\begin{equation}
\frac{\vnorm{\Yv}^2}{2 \Lambda} 
=  \frac{1}{2 \Lambda} \sum_{k=1}^{2 \Lambda} Y^2_k  \, .
\end{equation}
The LLN gives
\begin{equation}
\label{LLN2}
\frac{\vnorm{\Yv}^2}{2 \Lambda} \pto \frac{1}{2 \Lambda} \sum_{k=1}^{\Lambda} 2 \sigma_k^2 
\qquad \text{as } \Lambda \to \infty
\end{equation}
where
\begin{equation}
\label{Ybound}
\frac{1}{1 + \alpha \itp/\Lambda}  < \frac{1}{2 \Lambda}  \sum_{k=1}^{\Lambda} 2 \sigma_k^2 < 1
\end{equation}
as a result of \eqref{sigma_bound}. The convergence in probability in \eqref{LLN2} is guaranteed because $\var \left[ {\vnorm{\Yv}^2}/{2 \Lambda} \right]$ is finite as $\Lambda \to \infty$ (as will be verified in the next section) \cite{Varadhan2001}. Hence, the pre-factor $\sqrt{\Lambda} / \vnorm{\Yv}$ in \eqref{uL2} converges to a finite limit, implying that $\uL$ behaves asymptotically like a normal random variable. 

What is the variance? The variance of $\uL$ can be computed by brute force (i.e.~summing the variance $\sigma_k^2$ of each random variable $Y_k$) or by simply recalling the normalization \eqref{Econd} and invoking Lemma \ref{space_ind} (spatial independence) to get $\var \left( \uL \right) = 1/2$ for any $\Lambda$. Therefore,
\begin{equation}
\uL \sim \NN(0, 1/2)
\quad \text{approximately as } \Lambda \to \infty
\end{equation}
where the sense of convergence will be made more precise in the next section.
\end{proof}

\begin{proof}[{\bf Rigorous proof}]

The preceding argument gives the main intuition underlying Theorem \ref{thm2}, but it is not rigorous since the naive substitution $\vnorm{\xv}^2 \approx 2\Lambda / \alpha$ that produced  \eqref{LinGibbsApprox} was not justified. We will now determine the specific value of $\alpha$ for which this substitution can be justified, and we will bound the error that is incurred. The key is to perform the change of measure to $d \gssS$ while controlling the integrand carefully. We break the proof into a few parts, the first of which is a domain truncation that will eventually enable the change of measure.

\vsp{1}\noindent{\bf Domain truncation:}
Consider a random vector $\Xv \sim \NN(0,\Id)$, and the associated random variable
\begin{equation}
\Xsb = \frac{1}{2 \Lambda} \sum_{k=1}^{2\Lambda} X_k^2
=   \frac{\vnorm{\Xv}^2}{2\Lambda}
\end{equation}
The random variable $\Xsb$ has an expected value $\EV \left[ \Xsb \right] = 1$ and a variance of
\begin{equation}
\var \left( \overline{X^2_{2 \Lambda}} \right) = \frac{1}{4 \Lambda^2} \sum_{k=1}^{2\Lambda} \var \left(X_k^2\right) = \frac{1}{ \Lambda}
\end{equation}
since $\var \left(X_k^2\right) = 2$ for each $k$.
Consider the set
\begin{equation}
A = \left\{ \abs{ \Xsb - 1 } < \Lambda^{-p} \right\}
\end{equation}
where $p>0$ will be chosen later. 
Applying Chebyshev's inequality  \cite{Varadhan2001} with $\epsilon = \Lambda^{-p}$ gives
\begin{equation}
\label{Cheby1}
\mathbb{P}[A^c] = \mathbb{P} \left\{ \abs{ \Xsb- 1 } \ge \Lambda^{-p} \right\}
\le  \Lambda^{2p} \, \var \left( \Xsb \right) =  {\Lambda}^{-1+2p} 
\end{equation}
which is equivalent to the measure bound $\gss(A^c) \le \Lambda^{-1+2p}$. To make the measure of $A^c$ asymptotically small, it suffices to chose any $p$ in the range $0 < p < 1/2$.

	We will now bound the integrand in \eqref{LinGibbs} in order to show convergence as $\Lambda \to \infty$. In particular, the normalization constant $\Zpart$ must be included in these bounds. First, the bound
\begin{equation}
\frac{1}{ \vnorm{\xv}^2 } \sum_{k=1}^{\Lambda} 
\frac{k^2}{\Lambda^2} \left(x_k^2 + x_{\Lambda+k}^2 \right)  < 1
\end{equation}
immediately gives
\begin{equation}
\exp \left( - \frac{\itp}{ \Lambda^2 \vnorm{\xv}^2 } 
\sum_{k=1}^{\Lambda} k^2 \left(x_k^2 + x_{\Lambda+k}^2 \right) \right)
\ge \exp \left( -\itp \right)
\end{equation}
Therefore, the normalization constant $\Zpart$ is bounded below by
\begin{equation}
\label{Zbound}
\Zpart := \int_{\RR^{2 \Lambda}} \exp \left( - \frac{ \itp}{ \Lambda^2 \vnorm{\xv}^2 } 
\sum_{k=1}^{\Lambda} k^2 \left(x_k^2 + x_{\Lambda+k}^2 \right) \right) \, d\gss
\ge \exp \left( -\itp \right)
\end{equation}
since $\gss$ is a probability measure.

Now, the argument of the exponential in \eqref{LinGibbs} is non-positive, and $f \in C_b(\RR \to \RR)$ is bounded $\abs{f} < M$. Therefore, \eqref{Zbound} gives an overall bound on the integrand in \eqref{LinGibbs},
\begin{equation}
\abs{ \Zpart^{-1} f(\UL(\xhat)) \exp \left( - \frac{ \itp}{ \Lambda^2 \vnorm{\xv}^2 } 
\sum_{k=1}^{\Lambda} k^2 \left(x_k^2 + x_{\Lambda+k}^2 \right) \right) }
\le M \exp \left( \itp \right)
\end{equation}
Importantly, this bound is independent of $\Lambda$, since by assumption $\itp = \oo(1)$. The bound \eqref{Cheby1}, shows that $\gss(A^c) \to 0$ as $\Lambda \to \infty$, and therefore
\begin{equation}
\int_{A^c} \Zpart^{-1} \, f(\UL(\xhat)) \, \exp \left( - \frac{ \itp}{ \Lambda^2 \vnorm{\xv}^2 } 
\sum_{k=1}^{\Lambda} k^2 \left(x_k^2 + x_{\Lambda+k}^2 \right) \right) \, d\gss \to 0
\quad \text{as } \Lambda \to \infty
\end{equation}
Of course, the same holds for any subset of $A^c$. In particular, taking a radius of $R^2 = 2\Lambda(1 + \Lambda^{-p})$ gives the containment $\left\{ \vnorm{\xv}^2 \ge R^2 \right\} \subset A^c$, and thus
\edit{
\begin{align}
\label{TruncDom}
\EV f(\uL) - \int_{\vnorm{\xv} < R} \Zpart^{-1} f(\UL(\xhat)) \,
\exp \left( - \frac{ \itp}{ \Lambda^2 \vnorm{\xv}^2 } 
\sum_{k=1}^{\Lambda} k^2 \left(x_k^2 + x_{\Lambda+k}^2 \right) \right) \, d\gss(\xv) \,     \to 0
\quad \text{as } \Lambda \to \infty \, .
\end{align}
}

\vsp{1}\noindent{\bf Change of measure:}
So far, we have successfully truncated the integration \eqref{LinGibbs} to a finite domain and shown that the expected value $\EV f$ is unaffected asymptotically as $\Lambda \to \infty$. The next step is to perform a change of measure to the anistropic $d \gssS$. First note that
\begin{equation}
\frac{1}{\vnorm{\xv}^2} = \frac{\alpha}{2 \Lambda} 
+ \frac{1}{2 \Lambda} \left( \frac{2\Lambda}{\vnorm{\xv}^2} - \alpha \right)
\end{equation}
where $\alpha > 0$. Therefore, \eqref{TruncDom} can be rewritten as
\edit{
\begin{equation}
\label{LinGibbs2}
\EV f(\uL) - \int_{\vnorm{\yv} < R}  \Zmod^{-1}  f(\UL(\yhat)) \,
\exp \left( - \frac{ \itp}{ 2\Lambda^3 } \left( \frac{2 \Lambda}{\vnorm{\yv}^2} - \alpha \right) 
\sum_{k=1}^{\Lambda} k^2 \left(y_k^2 + y_{\Lambda+k}^2 \right) \right) \, d\gssS(\yv) \,  \to 0
\quad \text{as } \Lambda \to \infty \, .
\end{equation}
}
Above, $d\gssS$ is the anisotropic Gaussian measure given by \eqref{dgSig}, and $\Zmod$ is a modified partition function that can be calculated explicitly in terms of $\Zpart$ if desired. The variables $\xv$ and $\yv$ are simply dummy integration variables, and we choose to write \eqref{LinGibbs2} in terms of $\yv$ to distinguish the notation for the random variables that will be introduced soon.

Now, the argument of the exponential in \eqref{LinGibbs2} is no longer non-positive over $\RR^{2\Lambda}$. In particular, it can grow arbitrarily large in the far-field, $\vnorm{\yv} \gg 1$, which is problematic for bounding the integrand. However, on the truncated domain, $\vnorm{\yv} < R$, the argument can be bounded as follows
\begin{equation}
- \frac{ \itp}{ 2\Lambda^3 } \left( \frac{2 \Lambda}{\vnorm{\yv}^2} - \alpha \right)
\sum_{k=1}^{\Lambda} k^2 \left(y_k^2 + y_{\Lambda+k}^2 \right) \le
\frac{\alpha \itp}{2 \Lambda} \sum_{k=1}^{\Lambda} 
\frac{k^2}{\Lambda^2} \left(y_k^2 + y_{\Lambda+k}^2 \right) \le
\frac{\alpha \itp}{2 \Lambda} \vnorm{\yv}^2 \le \alpha \itp \left( 1 + \Lambda^{-p} \right)
\end{equation}
Additionally, the normalization constant $\Zmod^{-1} \le \exp(\itp)$ can be bounded by the exact same reasoning as before, owing to the fact that $d \gssS$ is a probability measure. These results give the overall bound on the integrand in \eqref{LinGibbs2} 
\begin{equation}
\label{intbound}
\abs{ \Zmod^{-1} f(\UL(\yhat)) \, 
\exp \left( - \frac{ \itp}{ 2\Lambda^3 } \left( \frac{2 \Lambda}{\vnorm{\yv}^2} - 1\right) 
\sum_{k=1}^{\Lambda} k^2 \left(y_k^2 + y_{\Lambda+k}^2 \right) \right) }
\le M \exp \left( \itp + 2 \alpha \itp \right)
\end{equation}
which holds within the truncated domain $\vnorm{\yv} < R$. Importantly, this bound is independent of $\Lambda$ by the scaling assumption \eqref{scaling}.

We now apply the Chebyshev-inequality argument used before, except this time with the random variable $\Yv \sim \NN(0, \Sig)$ corresponding to the measure $d \gssS$. Consider the associated random variable
\begin{equation}
\Ysb = \frac{1}{2 \Lambda} \sum_{k=1}^{2\Lambda} Y_k^2
= \frac{\vnorm{\Yv}^2}{2\Lambda}
\end{equation}
This variable has expected value
\begin{equation}
\label{mu}
\mu := \EV \left[ \Ysb \right] = \frac{1}{2 \Lambda} \sum_{k=1}^{\Lambda} 2 \sigma_k^2 < 1
\end{equation}
The inequality above relies on the fact that $\sigma_k^2 < 1$ for all $k$ as expressed in \eqref{sigma_bound}. Meanwhile, the variance is given by
\begin{equation}
\label{varbound}
\var \left[ \Ysb \right] = \frac{1}{4 \Lambda^2} \sum_{k=1}^{2 \Lambda} \var \left[ Y_k^2 \right]
= \frac{1}{4 \Lambda^2} \sum_{k=1}^{\Lambda} 4 \sigma_k^4 < \frac{1}{\Lambda}
\end{equation}
which relies on the fact that $\var[X^2] = 2 \sigma^4$ for a normal random variable with variance $\sigma^2$.

Now consider the set
\begin{equation}
B = \left\{ \vnorm{\Ysb - \mu} < \Lambda^{-q} \right\}
\end{equation}
for some $q >0$ to be chosen later. Chebyshev's inequality gives
\begin{equation}
\mathbb{P} \left[ B^c \right] \le \Lambda^{2q} \, \var \left[ \Ysb \right] < \Lambda^{-1+2q}
\end{equation}
where the bound \eqref{varbound} has been applied. As before, this bound in probability is equivalently to the bound in measure $\gssS(B^c) < \Lambda^{-1+2q}$, suggesting that $q$ should be chosen in the range $0<q<1/2$. Furthermore, if $q \ge p$, then $\mu < 1$ implies the containment $B \subset \{ \vnorm{\yv} < R \}$. In that case, \eqref{intbound} provides a bound for the integrand in \eqref{LinGibbs2}, and $\gssS(B^c) \to 0$ as $\Lambda \to \infty$, which implies
\edit{
\begin{align}
\label{LinGibbs3}
\EV f(\uL) - \int_{B} \Zmod^{-1} f(\UL(\yhat)) \,
\exp \left( - \frac{ \itp}{ 2\Lambda^3 } \left( \frac{2 \Lambda}{\vnorm{\yv}^2} - \alpha \right) 
\sum_{k=1}^{\Lambda} k^2 \left(y_k^2 + y_{\Lambda+k}^2 \right) \right) \, d\gssS(\yv) \,   \to 0
\quad \text{as } \Lambda \to \infty \, .
\end{align}
}

Next, we would like to choose the degree of freedom $\alpha$ to make the argument of the exponential in \eqref{LinGibbs3} small on the set $B$. Recall that, on the set $B$,
\begin{equation}
\label{bound1}
\abs{ \frac{\vnorm{\yv}^2}{2 \Lambda} - \mu} < \Lambda^{-q}
\end{equation}
Comparing \eqref{LinGibbs3} and \eqref{bound1} makes it clear that \edit{$\alpha = 1/\mu$} should be chosen, which implies, through \eqref{mu} and \eqref{sigma}, that $\alpha$ is a root of the nonlinear function
\edit{
\begin{equation}
\label{alg}
F(\alpha) := 1 - \frac{\alpha}{\Lambda} \sum_{k=1}^{\Lambda} \frac{1}{1 + \alpha \itp k^2 / \Lambda^3}
\end{equation}
We note that $F(0)=1$ and $\lim_{\alpha \to \infty} F(\alpha) = 1 - \Lambda^3(\Lambda+1)(2\Lambda+1)/(6 \itp)$. The latter formula implies $\lim_{\alpha \to \infty} F(\alpha) < 0$ for sufficiently large $\Lambda$ as long as the scaling law $\itp = O(\Lambda^0)$ holds, in which case the intermediate value theorem guarantees a root to exist on the interval $0< \alpha < \infty$.}

Setting $\alpha$ equal to the root of \eqref{alg}, implies that
\begin{equation}
\frac{2 \Lambda}{\vnorm{\yv}^2} - \alpha = \oo(\Lambda^{-q})
\end{equation}
holds on the set $B$. Furthermore,
\begin{equation}
\frac{ 1}{ 2\Lambda^3 } \sum_{k=1}^{\Lambda} k^2 \left(y_k^2 + y_{\Lambda+k}^2 \right)
\le \frac{1}{2 \Lambda} \sum_{k=1}^{\Lambda} \frac{k^2}{\Lambda^2} \left(y_k^2 + y_{\Lambda+k}^2 \right) \le \frac{\vnorm{\yv}^2}{2 \Lambda}
\end{equation}
and the right-hand side is bounded on the set $B$. Therefore, on the set $B$, the argument of the exponential in \eqref{LinGibbs3} is $\oo(\Lambda^{-q})$, which simplifies \eqref{LinGibbs3} to
\edit{
\begin{align}
\EV f(\uL) - \int_{B} \Zmod^{-1} f(\UL(\yhat)) \, \left( 1 + \oo(\Lambda^{-q}) \right) \, d\gssS(\yv) \, \to 0
\quad \text{as } \Lambda \to \infty \, .
\end{align}
}
Finally, since $\gssS(B^c) \to 0$ as $\Lambda \to \infty$ and the integrand in the above is bounded, the domain of integration can once again be expanded,
\edit{
\begin{align}
\label{LinGibbs4}
\EV f(\uL) - \int_{\RR^{2 \Lambda}} f(\UL(\yhat)) \, d\gssS(\yv) \,  \to 0
\quad \text{as } \Lambda \to \infty \, .
\end{align}}
where the vanishing $\oo(\Lambda^{-q})$ contribution has been removed. 
In \eqref{LinGibbs4}, $\Zmod$ has been replaced by its limiting value $\Zmod \to 1$ as $\Lambda \to \infty$, which can be seen by simply taking $\phi \equiv 1$ as the test function. The only requirements on $p$ and $q$ are that $0< p \le q < 1/2$, so it suffices to choose $p = q = 1/3$ for example.

	Importantly, the measure $\gssS$ in \eqref{LinGibbs4} is Gaussian. Therefore, the steps between \eqref{uL2}--\eqref{LLN2} in the non-rigorous proof apply, with \eqref{varbound} providing the justification for \eqref{LLN2}. If $Z = \UL(\Yhat)$, where $\Yv \sim \NN(0, \Sig)$, then $Z$ converges in probability to a normal random variable by the LLN. Furthermore, \eqref{LinGibbs4} shows that $\EV f(\uL) - \EV f(Z) \to 0$ as $\Lambda \to \infty$, hence $\uL$ converges in distribution to a normal random variable.
The variance can be obtained by recalling the normalization \eqref{Econd} and invoking spatial independence, Lemma \ref{space_ind}. We have therefore established the desired convergence in distribution:
\begin{equation}
\edit{\uL \dto \NN(0, 1/2)}
\quad \text{as } \Lambda \to \infty
\end{equation}
\end{proof}

\subsection{An improved importance distribution}
\label{ImportanceSampling}

Central to the proof of Theorem \ref{thm2} is the construction of the approximate measure $\gssS$ from \eqref{dgSig}, with standard deviations $\sigma_k^2$ given by \eqref{sigma} and $\alpha$ a root of \eqref{alg}. Not only is this measure useful as a theoretical tool, but it may pave the way for new numerical algorithms designed to efficiently sample from the Gibbs ensemble \eqref{Gibbs2}.

\begin{figure}
\begin{center}
\includegraphics[width = 0.89 \linewidth]{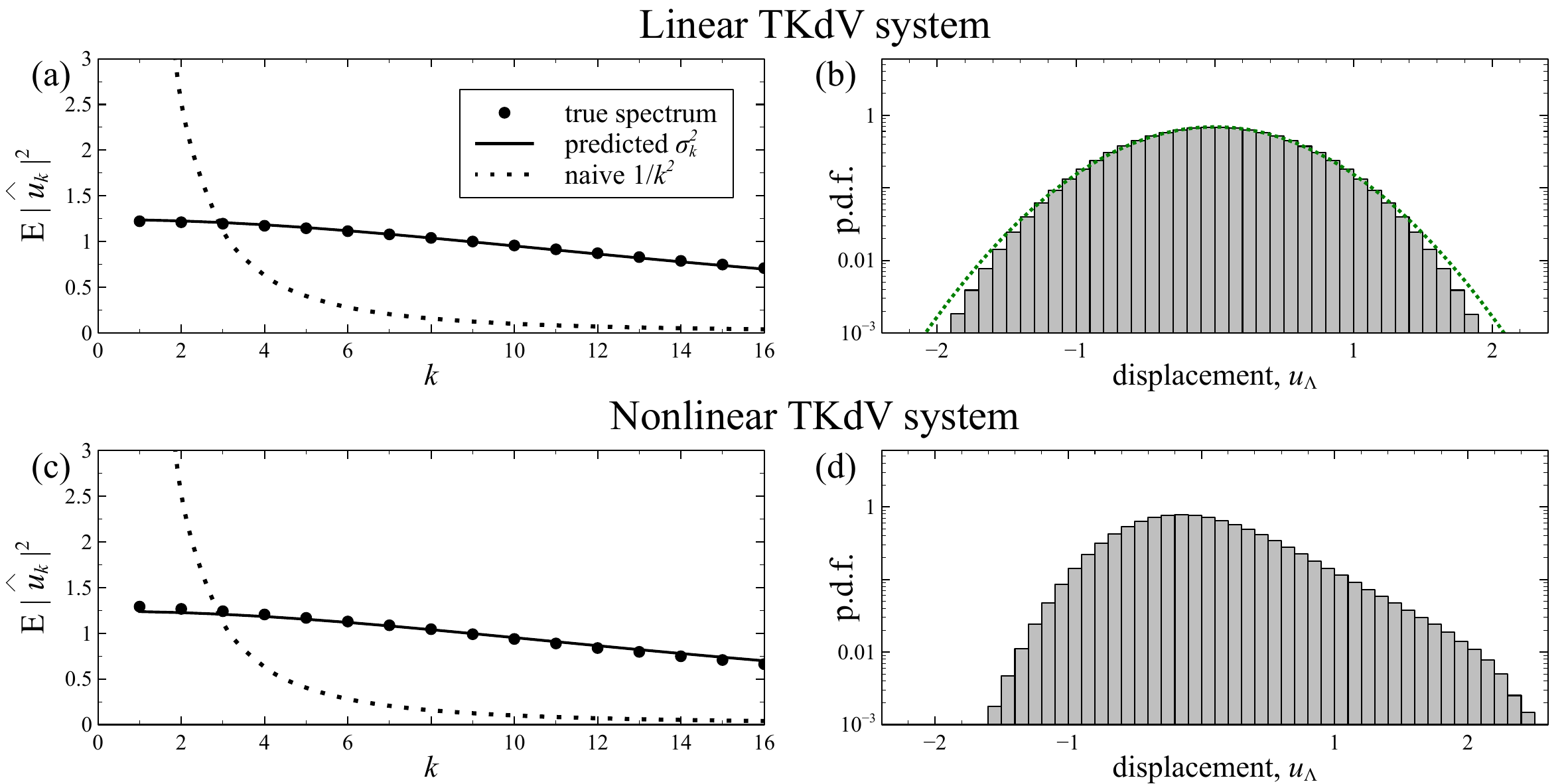}
\caption{
\edit{ Spectra and histograms for linear TKdV (top) and nonlinear TKdV (bottom) with $\Lambda = 16$ and $\itp = 10$. (a) The numerically computed spectral density, $\EV \abs{\uhat_k}^2$ (dots), is well predicted by the variances $\sigma_k^2$ (solid) from Eq.~\eqref{sigma}. The dotted curve shows the naive Weiner-measure decay rate, $1/k^2$, which is not a good approximation to the true spectrum. (b) The surface-displacement histogram shows nearly Gaussian statistics for the linear system.
(c) For nonlinear TKdV with $C_3/C_2 = 600$ the spectrum is still well predicted by Eq.~\eqref{sigma}. (d) The histogram, however, shows a sensitive response to nonlinearity and exhibits a strong positive skewness.}}
\label{Fig_bp10}
\end{center}
\end{figure}

\edit{
In particular, preliminary numerical tests indicate that the projection of $\gssS$ onto the hypersphere, $\En =1$, accurately captures the spectral decay of Gibbs measure $d\Gibbs$ from \eqref{Gibbs2}. Figure \ref{Fig_bp10}(a) illustrates this idea in the case of linear TKdV ($C_3=0$), with $\Lambda = 16$ and $\itp = 10$. In this figure, the true spectrum of $d\Gibbs$, shown by the dots, is obtained numerically by the sampling-importance resampling (SIR) algorithm \cite{skare2003improved}, where the instrumental distribution is taken to be the uniform measure $d\Gibbs_0$ from \eqref{Gibbs0}. 
That is, microstates $\{\uhat_k , k=1,2,\cdots,\Lambda \}$ are sampled from $d\Gibbs_0$, then resampled with probability determined by their Hamiltonian.
Roughly $10^7$ microstate samples are used. This is essentially a brute-force computation requiring significant resources in CPU time and memory. Meanwhile, Fig.~\ref{Fig_bp10}(a) shows that the simple formula \eqref{sigma} for $\sigma_k^2$ (solid curve) accurately predicts the decay structure, with the only numerical requirement being the computation of $\alpha$ as a root of \eqref{alg}. This close agreement suggests that $\gssS$ may serve as an improved instrumental distribution that samples the important regions of $d\Gibbs$ more faithfully than the uniform measure. For comparison, we also show the naive prediction $1/k^2$ associated with Weiner measure (dotted curve), which clearly disagrees with the true spectrum. This simple decay structure, and the accompanied convenience of working with Weiner measure, applies only to the bounded-energy formalism, $\En<\En_0$, used in other works \cite{Lebowitz1988, Lebowitz1989, Bourgain1994, Bourgain1999}. In the figure, all spectra have been normalized to have unit mean so as to facilitate comparison.
}

\edit{
	For this case of {\em linear} TKdV, Theorem \ref{thm2} suggests that the corresponding microstate statistics should be nearly Gaussian, provided that the scaling law on $\invtemp$ is reasonably satisfied. To offer numerical confirmation of this important fact, Fig.~\ref{Fig_bp10}(b) shows a histogram of surface displacements, $\uL$, obtained by inverse discrete Fourier transform of each sample $\{\uhat_k , k=1,2,\cdots,\Lambda \}$. The histogram, shown on a log-scale, is seen to agree closely with a normal distribution of the same variance (green dotted curve).

How does the picture change when nonlinearity is reintroduced? Figure \ref{Fig_bp10}(c)--(d) shows the corresponding spectra and histogram for the same set of parameters ($\Lambda=16$, $\itp=10$), but with $C_3/C_2=600$ so as to introduce nonlinearity into the system (due to the normalization of the inverse temperature, only the ratio $C_3/C_2$ needs to be specified).
Interestingly, the spectral decay seen in Fig.~\ref{Fig_bp10}(c) is nearly unchanged and remains well-predicted by formula \eqref{sigma}. Evidently, the spectrum is controlled primarily by the contribution $\Htwo$ associated with the linear term, and is relatively insensitive to the contribution $\Hthree$ from the nonlinear term.
Moreover, the decay rate of the spectrum is rather gradual, much more gradual than the $1/k^2$ Weiner-measure decay for instance. This observation sheds light on previous dynamical simulations of KdV, which, when run until equilibrium, produced a broad spectrum with a gradual decay rate (up until the sudden drop-off near the largest resolved wave-numbers due to numerical truncation) \cite{pelinovsky2006numerical, Majda2019, MajdaQi2019, MooreJNS2020}. A similar gradual decay of the spectrum has been observed in statistical and dynamical studies of the truncated Burgers-Hopf system \cite{Abramov2003, Bajars2013}.

While the spectrum is relatively insensitive to nonlinearity, the histogram shown in Fig.~\ref{Fig_bp10}(d) displays a visible response. What was previously a symmetric histogram has become conspicuously skewed in the direction of positive $\uL$. The slower decay of the tail as $\uL \to +\infty$ corresponds to events of large surface displacement, e.g.~rogue waves, occurring on a more frequent basis, as has been borne out numerically by other studies \cite{Majda2019, MajdaQi2019, MooreJNS2020}.
}

	The fact that the spectrum remains well approximated by the variances $\sigma_k^2$ calculated from formulas \eqref{sigma} and \eqref{alg}, even with the reintroduction of nonlinearity, suggests that the measure $\gssS$ from \eqref{dgSig} could serve as the foundation for a highly efficient importance sampling algorithm \cite{Owen2013}. In particular, this measure generates microstates with the proper spectral decay that could then be input into an acceptance-rejection step to generate {\em independent} samples from the Gibbs ensemble. Roughly speaking, since the spectral decay has already been accounted for, there would only remain one criterion for acceptance, namely sufficient skewness. This would suggest a relatively high acceptance rate. Unlike the commonly used Markov-Chain Monte Carlo (MCMC), which produces correlated samples and does not have good parallelization properties, the proposed algorithm would generate {\em independent} samples and would be easy to parallelize. Though beyond the scope of the current paper, this is an exciting avenue for future work.

\begin{figure}
\begin{center}
\includegraphics[width = 0.89 \linewidth]{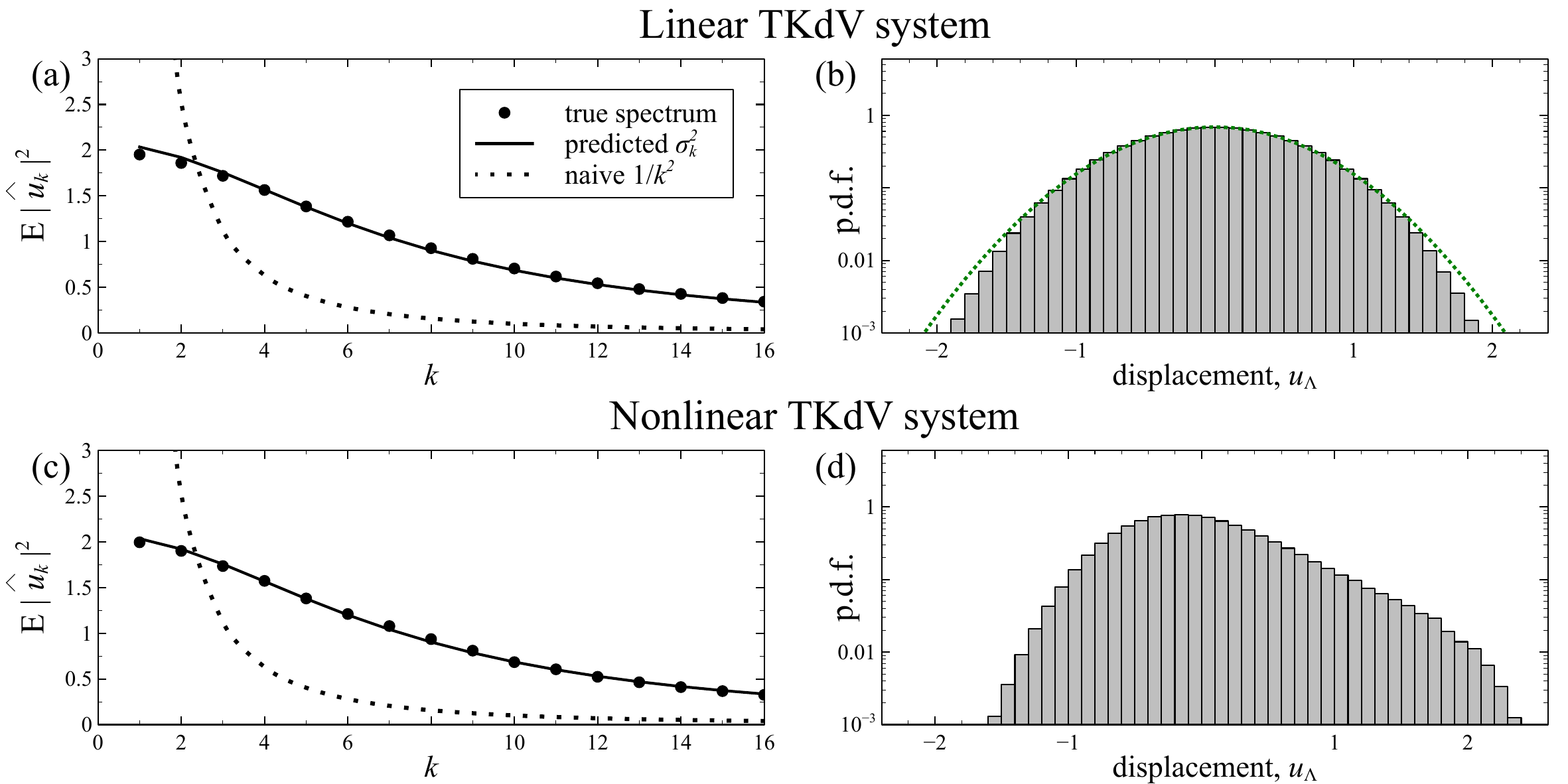}
\caption{
\edit{Application of the theory to the experimental parameters used by Moore et al.~\cite{MooreJNS2020}. Even with a value of $\itp = 40$ that lies outside the region of asymptotic validity, the formula \eqref{sigma} still captures the proper spectral decay for both linear (a) and nonlinear (c) TKdV. (b) The linear system exhibits nearly Gaussian microstate statistics. (d) The nonlinear system, with value $C_3/C_2 = 140$ taken from laboratory experiments, shows a strong positive skewness as is consistent with the experimental measurements. In all cases the number of modes is $\Lambda = 16$.
}}
\label{Fig_bp40}
\end{center}
\end{figure}

\edit{
The parameters in the above numerical example, $\Lambda = 16$ and $\itp = 10$, were chosen to showcase the accuracy of our theory when the scaling law \eqref{scaling} is reasonably satisfied. A natural question is whether these results extend to realistic experimental parameters. To address this question, we now test the theory with the experimental parameters used by Moore et al.~\cite{MooreJNS2020}. In particular, direct measurements of the wave amplitude, along with input parameters such as the water depth and peak forcing frequency,  yield the ratio $C_3/C_2 = 140$. Meanwhile, by matching the surface-displacement skewness observed downstream, Moore et al.~\cite{MooreJNS2020} was able estimate the system's inverse temperature. With $\Lambda=16$ modes, this estimate produces, in the notation of the current paper, the value $\itp = 40$. Interestingly, this value does not obviously satisfy the scaling condition $\itp = O(\Lambda^0)$, thus providing a challenging test for our theory.
Figure \ref{Fig_bp40}(a)--(b) shows the spectrum and surface-displacement histogram for the linear system ($C_3 =0$) with $\Lambda=16$ and $\itp = 40$. Somewhat surprisingly, the formula \eqref{sigma} still captures the spectral decay accurately, even with this relatively large value of $\itp$. Figure \ref{Fig_bp40}(b) shows that the surface-displacement statistics are nearly Gaussian in this linear case. Meanwhile, Figure \ref{Fig_bp40}(c)--(d) shows the spectrum and histogram for the nonlinear system, with the true experimental parameters of $C_3/C_2 = 140$ and $\itp = 40$. Even with the reintroduction of nonlinearity and the large value of $\itp$, the formula \eqref{sigma} still approximates the numerically-sampled spectrum well. Thus, formula \eqref{sigma} seems to provide a good approximation even outside the strict asymptotic regime in which it was derived. Figure \ref{Fig_bp40}(d) shows that the strength of nonlinearity is sufficient to severely skew the surface-displacement statistics, as is consistent with the experimental measurements of Moore et al.~\cite{MooreJNS2020} .
}

\section{Conclusions}
\label{Conclusion}

	This paper reports two fundamental results, Theorems \ref{thm1} and \ref{thm2}, on the surface-displacement distributions implied by a canonical-microcanonical Gibbs ensemble of the truncated KdV system. Theorem \ref{thm1} establishes that vanishing inverse temperature, $\invtemp=0$, implies Gaussian displacement statistics in the limit of large cutoff-wavenumber, $\Lambda \to \infty$. 
\edit{
For $\beta >0$, our numerical sampling experiments demonstrate that the linear TKdV system can produce symmetric, yet highly non-Gaussian, statistics, contrasting with the common intuition that linearity leads to a Gaussian state. To offer resolution, we identify a precise scaling law, $\invtemp C_2 = \oo(\Lambda^{-2})$, that guarantees convergence to Gaussian statistics in the linear system, as established by Theorem \ref{thm2}. 
In particular, the fixed-energy formalism, which has been recognized by others as the most physically relevant \cite{Lebowitz1988, Lebowitz1989}, renders the convergence to Gaussian non-trivial. The scaling law clearly delineates two parameter regimes: that in which the intuitive association of linearity with Gaussian statistics holds, and that in which it does not. 
Taken together, the new rigorous results imply necessary conditions for anomalous wave behavior, namely non-zero inverse temperature {\em and} either nonlinearity or the violation of scaling law \eqref{scaling}.
}

\edit{
In the regime of stronger nonlinearity, the form of $\Hthree$ in \eqref{H3H2} shows that nonlinearity promotes skewed microstates in the Gibbs ensemble \eqref{Gibbs2}, in agreement with both laboratory and numerical experiments. However, determining the precise skewed distribution to which surface-displacement statistics converge under those conditions remains an open question. One candidate suggested by empirical evidence is a mean-zero gamma distribution \cite{Bolles2019, Majda2019}. In future work, we hope to extend the foundation established here to examine such possibilities.
}

	In addition to the end results, the proofs of Theorems \ref{thm1} and \ref{thm2} provide important insights into TKdV statistical mechanics that will likely prove valuable in future studies. First, it is a remarkable coincidence that the same scaling relationship, $\invtemp C_2 = \oo(\Lambda^{-2})$, obtained from physical reasoning in earlier work \cite{MooreJNS2020} serves a central role in the proof of Theorem \ref{thm2}. Without this relationship, the argument of the exponential in \eqref{LinGibbs} cannot be controlled, thus making it difficult to bound the error incurred by the change of measure. Given this coincidence, it is possible that the scaling law \eqref{scaling} plays a more fundamental role in the statistical mechanics of TKdV and KdV than has been previously recognized. In particular, if one were to extend similar analysis to the continuous KdV system with non-vanishing dispersion, the scaling law \eqref{scaling} suggests taking $\invtemp \to 0$ at a particular rate. There exist complementary results from statistical-mechanics analysis of continuous PDE systems \cite{Randoux2014, Costa2014, Agafontsev2015}  versus discrete counterparts \cite{Abramov2003, MajdaWang2006, MajdaQi2019}, but it is not always clear how to reconcile the two. For the case of KdV, perhaps scaling law \eqref{scaling} can bridge these different perspectives.
	
	Lastly, our proof of Theorem \ref{thm2} relies on the construction of an approximating measure $\gssS$, given in \eqref{dgSig}. For linear TKdV, we proved that the projection of this measure onto the hypersphere approximates the Gibbs measure $d\Gibbs$ as $\Lambda \to \infty$. Moreover, numerical experiments suggest this measure to accurately capture the spectral decay rate of $d\Gibbs$, even upon the reintroduction of nonlinearity. This observation suggests that $\gssS$ may serve as the foundation for an efficient and easily parallelized importance sampling algorithm to generate {\em independent} samples of the Gibbs ensemble. This possibility will be explored in future work.
	
\section{Acknowledgements}
N.J.~Moore would like to acknowledge support from the National Science Foundation, DMS-2012560, and from the Simons Foundation, award 524259.
	
\bibliographystyle{plain}
\bibliography{wavesbib}

\end{document}